\documentclass[fleqn,11pt]{article}

\usepackage{amsfonts,amsmath,amssymb}
\usepackage{latexsym}

\usepackage{amsfonts,amssymb,cite}
\usepackage{graphicx}

\def\noi{\noindent}
\def\jnumber#1#2{\thispagestyle{empty} \noi\unitlength=1mm
    	\begin{picture}(178,10)
            \put(177,15){\llap{\large\it Grav. Cosmol. No.\,#1, #2}}
                    \end{picture}}

\newcommand{\Title}[1]{\noi {{\Large\bf #1}}\\[1ex]}

\def\Aunames#1{\noi{\bf #1}}
\def\au#1{${}^{#1}$}
\def\Addresses#1{\medskip\noi \protect
	\begin{description}\itemsep -3pt {\it #1} \end{description}}
\def\adr#1#2{\item[${}^{#1}$]{\it #2}}

\newcommand{\Abstract}[1]{\vskip 2mm \begin{center}
        \parbox{16.4cm}{\small\noi #1} \end{center}\medskip}

\def\email#1#2{\footnotetext[#1]{e-mail: #2}\addtocounter{footnote}{1}}

\def\nqq{\hspace*{-2em}}

\def\cm{\hspace*{1cm}}

\usepackage{color}

\def\Acknow#1{\subsection*{Acknowledgments} #1}
\def\Funding#1{\subsection*{Funding} #1}

\def\Jl#1#2{#1 {\bf #2},\ }

\def\ApJ#1 {\Jl{Astroph. J.}{#1}}
\def\CQG#1 {\Jl{Class. Quantum Grav.}{#1}}
\def\DAN#1 {\Jl{Dokl. AN SSSR}{#1}}
\def\GC#1 {\Jl{Grav. Cosmol.}{#1}}
\def\GRG#1 {\Jl{Gen. Rel. Grav.}{#1}}
\def\IJMPD#1 {\Jl{Int. J. Mod. Phys. D}{#1}}
\def\JETF#1 {\Jl{Zh. Eksp. Teor. Fiz.}{#1}}
\def\JETP#1 {\Jl{Sov. Phys. JETP}{#1}}
\def\JHEP#1 {\Jl{JHEP}{#1}}
\def\JMP#1 {\Jl{J. Math. Phys.}{#1}}
\def\NPB#1 {\Jl{Nucl. Phys. B}{#1}}
\def\NP#1 {\Jl{Nucl. Phys.}{#1}}
\def\PLA#1 {\Jl{Phys. Lett. A}{#1}}
\def\PLB#1 {\Jl{Phys. Lett. B}{#1}}
\def\PRD#1 {\Jl{Phys. Rev. D}{#1}}
\def\PRL#1 {\Jl{Phys. Rev. Lett.}{#1}}

\def\lal{&&\nqq {}}

\def\beq{\begin{equation}}
\def\eeq{\end{equation}}
\def\bear{\begin{eqnarray}}
\def\bearr{\begin{eqnarray} \lal}
\def\ear{\end{eqnarray}}
\def\earn{\nonumber \end{eqnarray}}

\def\nnn{\nonumber\\ \lal }

\begin{document}
\twocolumn[
\jnumber{issue}{year}

\Title{On particle collisions in the vicinity of the charged black holes}

\Aunames{Timur Pryadilin,\au{a,1} Daniil Zhitov,\au{b,2} and Vitalii Vertogradov\au{c,3}}

\Addresses{
\adr a {Department of Applied Mathematics and Theoretical Physics, University of Cambridge, CB3 0WA, UK}
\adr b {Cavendish Laboratory, University of Cambridge, Cambridge, CB3 0HE, UK}
\adr c {Physics department, Herzen state Pedagogical University of Russia,
		48 Moika Emb., Saint Petersburg 191186, Russia
		SPb branch of SAO RAS, 65 Pulkovskoe Rd, Saint Petersburg 196140, Russia}
}


\Abstract
	{The process of particle collision in the vicinity of black holes is known to generate unbounded energies in the center-of-mass frame (the Bañados-Silk-West (BSW) effect) under specific conditions. We consider this process in the charged black hole metrics, namely, the Reissner-Nordstrom (RN) and Majumdar-Papapetrou (MP) metrics. We consider the energy extraction from Bardeen regular black hole due to BSW effect. Like in RN case, we show that there is no restriction on energy extraction, but for real charged particles this effect is negligible. We derive necessary and sufficient conditions for this process. The conditions for the BSW effect in RN and MP metrics are shown to be identical, which is explained by the asymptotic equivalence of the two metrics near the horizons. Energy extraction in the RN metric is discussed. It is shown that if two real particles collide while falling onto a black hole, they are extremely unlikely to generate an ultra-massive particle. For the case of head-on collisions, we derive an upper bound on extracted mass, which depends on the lapse function of the metric at the point of collision.}
\medskip

] 
\email 1 {tp520@cam.ac.uk}
\email 2 {dz337@cam.ac.uk}
\email 3 {vdvertogradov@gmail.com\\ \cm (Corresponding author)}

\section{Introduction}

  Although we know about  solutions of the Einstein equations which describe black holes  for more than a century, only several years ago, the direct observation of gravitational waves~\cite{bib:ligo} and black hole shadow~\cite{bib:event} made them real astrophysical objects. Despite light can't escape the black hole, one can know its properties through the influence on the surrounding matter. Black holes can serve as an arena for high energy physics~\cite{bib:pavlov_complex}. 

Penrose~\cite{bib:pen} showed that there might be particles with negative energy in the ergosphere of a rotating black hole that can be used to extract its rotational energy. Another example are charged particles which move in charged black hole background~\cite{bib:chandra, bib:ruf}. Thereby, the Penrose-like process for charged black holes is also possible. This question has been studied for Reissner--Nordstrom~\cite{bib:zaslav} and Majumdar--Papapetrou~\cite{bib:richard} cases.

Ba\~nados, Silk, and West demonstrated that particles collision near the event horizon of extremal Kerr black hole~\cite{bib:bsw} can achieve arbitrarily high center-of-mass energy if the angular momentum of either of the incident particles is fine-tuned. The energy extraction is also possible due to this effect~\cite{bib:japan}. In Schwarzschild case, one can't get unbound center-of-mass energy due to this process. This process for naked singularity is investigated in papers~\cite{bib:joshi1, bib:joshi2}. However, in the Reissner--Nordstrom-anti-de-Sitter case~\cite{bib:zaslav2}, there is innermost stable equilibrium point and the center-of-mass energy of two particles collision in this point can be unbound. The energy extraction has been considered~\cite{bib:zaslav3}
and it has been shown that one can extract arbitrarily large amount of energy. The generalization of this process for charged Vaidya dynamical black hole has been studied in~\cite{bib:charged_vaidya}

It was shown that negative energy particles in the ergoregion of a rotating black hole, follow so-called white hole geodesics i.e. they appear in the ergoregion from the gravitational radius~\cite{bib:grib, bib:vertogradov}. The same is valid for trajectories of charged particles with negative energy in Reissner--Nordstrom~\cite{bib:zaslav} and partly in Majumdar--Papapetrou ~\cite{bib:richard} spacetimes. It means that one can consider a front collision near the event horizon, that can lead to unbound amount of energy in the centre of mass $E_{c.m.}$~\cite{bib:pavlov}. This process has been considered for a collapsing matter cloud ~\cite{bib:vaidya}.

In this paper, we consider the energy extraction from Bardeen regular black hole~\cite{bib:bardeen} due to BSW effect. We demonstrate that the energy extraction can be arbitrary large like in the case of an extremal Reissner-Nordstrom case~\cite{bib:zaslav3}. We consider the BSW effect and front collision for charged Reissner--Nordstrom black hole and system of two charged Majumdar--Papapetrou black holes~\cite{bib:hawking}. The energy extraction of these processes is also studied for real particle cases like an electron and proton and showed that for real particles the process leads to small amount of an extracting energy.

This paper is organized as follows. In sec.II we describe the methods of BSW effect and extrating energy in sec. III we consider the center-of-mass energy of two particles collision and the energy extraction in Reissner--Nordstrom  case. In sec. IV the same effects are considered for binary system of two charged black hole in Majumdar--Papapetrou case and in regular Bardeen black hole. Sec. V is the conclusion.

The system of units $c=G=1$ will be used throughout the paper. The signature is $-\,+\,+\,+$.

\section{Methods}
\subsection{Basic definitions}
We begin by formulating a general method for analyzing particle kinematics in different backgrounds. It is introduce it by applying to the Reissner--Nordstr\"om solution of Maxwell-Einstein theory, describing a static spherically-symmetric charged black hole:
\begin{equation}\label{metric_RN}
ds^2=-f dt^2 + \frac{1}{f} dr^2 + r^2d\Omega^2 \,,
\end{equation}
where the lapse function $f$ is given by:
\begin{equation}\label{lapse_RN}
f=1-\frac{2M}{r}+\frac{Q^2}{r^2}=\frac{(r-r_{-})(r-r_{+})}{r^2} \,,
\end{equation}
\begin{equation}\label{roots}
    r_{\pm}=M\pm\sqrt{M^2-Q^2},
\end{equation}
and the electrostatic potential is
\begin{equation}\label{potential_RN}
\varphi(r)=\frac{Q}{r} \,.
\end{equation}
Here $M$ is the mass of a black hole, $Q$ its electric charge, $r_{+}$ and $r_{-}$ represent the radial coordinates of the event and Cauchy horizons respectively. The black hole is extremal if $Q=M$, and if $|Q|>M$ the singularity is naked. The latter case is not considered in this article. Without loss of generality, we assume that $Q>0$.

In this background, motion of charged particles is non-geodesic, due to the electrostatic force. Still, the spherical symmetry implies the motion is planar. For convenience, we consider the equatorial plane $\theta=\frac{\pi}{2}$. Temporal and spherical symmetries also provide constants of motion $E$ and $L$ - energy and angular momentum, both defined per unit mass. We also specify $\gamma$,  charge-to-mass ratio of the particle. 
The equations of motion provide us the four-velocity $u^\mu=dx^\mu/d\tau$

\begin{align}\label{rn_eq_of_motion}
u^0&=\frac{dt}{d\tau}=f^{-1}\left(E-\frac{\gamma Q}{r}\right)\,, \\
u^1&=\frac{dr}{d\tau}=\pm \sqrt{\left(E-\frac{\gamma Q}{r}\right)^2-\left(\frac{L^2}{r^2}+1\right)f}\,, \\
u^2&=\frac{d\theta}{d\tau}=0\\
u^3&=\frac{d\varphi}{d\tau}=\frac{L}{r^2}\,.
\end{align}

The sign in the expression for $u^1$ is determined by whether the particle moves inward or outward. Moreover, a necessary condition is $u^0>0$ outside the event horizon (so-called "forward-in-time" condition).

In the BSW effect, there are two particles of the same mass $m_0$ with four-velocities $u_{(1)}^\mu$ and $u_{(2)}^\mu$. The collision energy measured in the center of mass frame is given by~\cite{bib:bsw}
\begin{equation}\label{ecm}
    \frac{E^2_{cm}}{2m^2}=1-g_{ik}u_1^i u_2^k \,.
\end{equation}

Using the equations \eqref{rn_eq_of_motion}, we get
\begin{equation}\label{ECM}
\frac{E^2_\text{cm}}{2m^2}-1=\frac{1}{f}\left(X_1X_2\pm\sqrt{X_1^2-Y_1f}\sqrt{X_2^2-Y_2f}\right)-\frac{L_1^2L_2^2}{r^2} \,.
\end{equation}

Here we have introduced new quantities:
\begin{equation}\label{def_xy}
X_{i}=E_{i}-\frac{\gamma_{i}r_Q}{r},\quad Y_{i}=\frac{L_{i}^2}{r^2}+1,\quad i=1,2\,.
\end{equation}

The physical interpretation of $X$ is that it behaves similarly to the kinetic energy in classical electromagnetism. We have from \eqref{rn_eq_of_motion} we see that $u^0=dt/d\tau=X/f$. Note, that the "forward-in-time" condition restricts accessible particle radii by the $X>0$ condition, that is similar to the positivity of classical kinetic energy. Also, the arbitrary sign denotes to essentially different types of collision, determined to the relative signs of $dr/d\tau$ of the two particles. We consider the two cases separately.
\subsection{Near-horizon expansions}
The later analysis requires certain care for detail, so we discuss approximations. A reliable technique is Taylor expanding various quantities in powers a of dimensionless quantity $\delta=(r-r_+)/r_+$.
    \bearr\label{X_taylor}
        X = E-\frac{\gamma Q}{r} = E - \frac{\gamma Q}{r_+}\left(1-\delta+\delta^2+...\right)
        \nnn
        =X_h +\frac{\gamma Q}{r_+}\delta - \frac{\gamma Q}{r_+}\delta^2+...
    \ear
    where $X_h=E-\frac{\gamma Q}{r_+}$ is the value of $X$ on the horizon. We also need to expand $X^2$:
    \begin{equation}\label{X^2_taylor}
        X^2 = X_h^2 +\frac{2\gamma QX_h}{r_+}\delta +\left(\left(\frac{\gamma Q}{r_+}\right)^2-\frac{2\gamma Q X_h}{r_+}\right){\delta^2}+...
    \end{equation}
    Providing an expansion for $f$ is slightly trickier due to the presence of another scale $(r_+-r_-)/r_+$. Then:
    \begin{equation}\label{f_noncrit_taylor}
        f=\alpha \delta+...,
    \end{equation}
    where $\alpha\geq0$ is a constant coefficient, exact value of which is irrelevant. The subtlety is that at distances much larger than $r_+-r_-$ the quadratic term would be larger than the linear one. Then, when one considers extremal black holes, for which $r_+=r_-$, the leading term is quadratic:
    \begin{equation}\label{f_crit_taylor}
        f={\delta^2}+...
    \end{equation}
    In practical terms, the condition of extremality that we impose later, is not quite that $r_+-r_-=0$, but that it is much smaller than the collision-horizon distance $r_+\delta$.
\subsection{Restrictions from particle kinematics}
This discussion already provides us with some useful information. The region accesible to a particle is determined by the conditions:
\begin{enumerate}
    \item $X > 0$ ("forward-in-time" outside the outer horizon)
    \item $X^2-Yf\geq 0$ (turning points)
\end{enumerate}
In the vicinity of the horizon, as long as $X_h$ finite, both conditions are easily satisfied. However, we would also be interested in "critical" particles, such that $X_h=0$. Then, since $X(r_+)=E-\gamma Q/r_+=X_h=0$, the condition 1 implies $\gamma>0$. Hence, a critical particle must have the same charge sign as a black hole.

In condition 2, in the vicinity of the horizon, the first two terms of the expansion vanish, and the leading term is quadratic in $\delta$. At the same time $f$ could be linear (with $\alpha>0$) for non-extremal holes. This means that critical particles cannot approach the vicinity of the horizon in this case due to violation of the second condition. If the hole is extremal ($r_-=r_+=Q=M$), the condition (to the leading order) transforms to $\gamma^2-\left(\frac{L^2}{r_+^2}+1\right)\geq 0$. Therefore, \emph{critical particles can achieve the horizon only if the hole is extremal}, and their angular momentum is bounded
\begin{equation}
|L|\le Q \sqrt{\gamma^2-1}\,.
\end{equation}
Of course, this also implies that $\gamma\geq1$.
 Moreover,  for an extremal black hole with $r_+=Q$, the condition of particle criticality gives $E_1=\gamma_1$. We thus conclude that a critical particle has $E_1\ge 1$, so, its trajectory always originates at infinity (except for a particular case $E_1=1$, where a particle is in a state of indifferent equilibrium at every point).

\subsection{Centre-of-mass collision energy}

Now let's analyse the expression for the collision energy \eqref{ECM}. The possibility of an infinite energy arises from the $1/f$ factor, which becomes unbounded in the vicinity of the event horizon (where $g_{00}=-f=0$).
\begin{itemize}
    \item "+" case: this corresponds to the particles moving in opposite directions.  $f\to 0$ next to the horizon, the expression in brackets tends to $2X_{1h}X_{2h}$, which is finite unless either particle is critical. Thus, infinite CoM energy arises for all kinds of particles, that are non-critical and can achieve the horizon.
    \item "-" case: this case is more nuanced, since the nominator tends to zero as well. 
    
    Finally, we can proceed to analyze the singular term of \eqref{ecm}. Assume $X_{1h}, X_{2h}$ are both finite. Then $X_1,X_2$ are also in the vicinity of the outer horizon. Then for collision at a point such that $f_c\ll X_{1c},X_{2c}$ (subscript 'c' for collision):
    \bearr\label{minus_case_expansion}
        \frac{1}{f_c}\left(X_{1c}X_{2c}-\sqrt{X_{1c}^2-Y_{1c}f_c}\sqrt{X_{2c}^2-Y_{2c}f_c}\right) 
        \nnn
        \approx\frac{1}{2}\left(\frac{X_{1c}Y_{2c}}{X_{2c}}+\frac{X_{2c}Y_{1c}}{X_{1c}}\right),
    \ear
    a result, essentially obtained before\cite{bib:zaslav4}. However, the intricacies of the approximation were not explicitly explained in the reference. One tends to think that this expression suggests that if one of the particles is critical $X_h\approx X_c=0$, infinite energy of collision is achievable. This is true, but is not so obvious: $X_c=0$ violates the condition $f_c\ll X_c$ implied in this derivation. In addition, as previously noted, a critical particle can only achieve the horizon if the hole is extremal.

    One could analyse it more thoroughly. Assume $X_{1h}=0$, $X_{2h}$ is finite. Expanding the expression in brackets the the first order of $\delta$, we obtain the singular contribution:
    \begin{equation}
        \approx\frac{1}{f}\left((\gamma_1-\sqrt{\gamma_1^2-1})X_{2h}\delta+O(\delta^2)\right)
    \end{equation}
    Given that $f=\delta^2 +O(\delta^3)$, the singular contribution to collision $E_{cm}^2/2m$ is 
    \begin{equation}
        \frac{(\gamma_1-\sqrt{\gamma_1^2-1})X_{2h}}{\delta}
    \end{equation}
\end{itemize}
We finish this section by writing the condition $\gamma \geq 1$ with SI units included:
\begin{equation}\label{el_ratio}
    \tilde{\gamma} = \frac{|q|}{m\sqrt{4\pi \varepsilon_0 G}}\geq 1,
\end{equation}
where $\varepsilon_0 \approx 8.854 \times 10^{-12}\,\mathrm{F}\,\mathrm{m}^{-1}$ is the vacuum permittivity, and $G\approx 6.674\times 10^{-11}\,\mathrm{N}\, \mathrm{m}^2\,\mathrm{kg}^{-2}$ is the gravitational constant. We call this dimensionless quantity $\tilde{\gamma}$ the \textit{electric ratio} of a particle. It is numerically equal to $|\gamma|$ in the units used previously.

\subsection{Energy extraction}

It was shown in~\cite{bib:zaslav3} that when the particles with opposite signs of radial momentum collide in the vicinity of a critical charged black hole, then the energy and mass of one of the out-coming particles admit lower, and not upper bounds. Namely, in a specific scenario when the observed particle moves towards the black hole, reaches a turning point just outside the horizon, and then flies to infinity, the following inequalities hold:
\begin{equation}\label{lower_bound}
    E \geq m \geq m_{\mathrm{min}} = |q_0| - \sqrt{q_0^2 - m_0^2},
\end{equation}
where $E$ and $m$ are energy and mass of the observed particle (in this section $E$ denotes absolute energy, and not energy per unit mass), and $q_0, m_0$ are charge and mass of the critical particle. Thus, the authors reach the conclusion of a possibility of creation of ultra-heavy particles detectable at infinity.

However, it is important to note that for real particles the lower bound is exceptionally small, which means that it is highly unlikely that the mass of the out-coming particle even exceeds the mass of the critical particle. Table \ref{tab:real_particles} shows the value of $A$ as well as the electric ratio $\tilde{\gamma}$ defined in for some subatomic particles in SI units, for which we used the formula
\begin{equation}
    m_{\mathrm{min}}=\frac{|q|}{\sqrt{4\pi\varepsilon_0 G}}+\sqrt{\frac{q^2}{4\pi\varepsilon_0 G}-m^2}.
\end{equation}

\begin{table*}[ht]
    \centering
    \begin{tabular}{p{2.5cm}|p{2cm}|p{2cm}|p{3.5cm}|p{3cm}}
         Particle &  Mass $m$/$m_e$ & Charge $q$/$e$ & Electric ratio $\tilde{\gamma}$&Mass bound $m_{\mathrm{min}}$/$m_e$ \\ \hline
         electron & $1.00\times 10^0$ & -1&$2.04\times10^{21}$ &$2.45\times 10^{-22}$\\
         proton & $1.84\times 10^3$& 1& $1.11\times 10^{18}$& $8.25\times 10^{-16}$\\
         $\alpha$-particle&$7.29\times 10^3$ & 2 &$5.60\times 10^{17}$ &$6.52\times 10^{-15}$\\
         ${}^{197}_{79}\mathrm{Au}$ & $3.59\times 10^5$& 79 &$4.49\times 10^{17}$ &$4.00\times 10^{-13}$
    \end{tabular}
    \caption{Lower bound of energy extraction for some subatomic particles}
    \label{tab:real_particles}
\end{table*}

As we can see from Table \ref{tab:real_particles}, all real subatomic particles satisfy $\tilde{\gamma}>1$ (in fact we can say that $\tilde{\gamma}\gg 1$), so they can in principle be critical particles. However, even for relatively heave particles like the nucleus of gold atom, the lower bound for the mass is many orders of magnitude lower than the mass of electron, so it only tells us that these processes cannot emit ultra-light particles like neutrinos. It is reasonable to assume that even if collisions with creation of ultra-heavy particles escaping to infinity do not violate conservation laws, and hence, are physically possible, they must be severely outnumbered by collisions that produce usual particles with masses smaller than the masses of collision's constituents.

We will now proceed to calculate energy extraction for the case when one of the particles is falling onto the black hole, and the other one is moving outward. We will ignore the question of how the latter particle could arise. As mentioned above, it follows from \eqref{ECM} that in this case the energy of collision is arbitrarily large when measured in the center-of-mass frame for all charged black holes (not necessarily extremal). Similarly, there are no longer any criticality conditions. Hence, we can expect a more significant energy extraction to be possible in this case.

As usual, we assume that there are 4 particles involved in the collision. Particles 1 and 2 collide to produce particles 3 and 4. We assume that particles 1 and 3 move outward, while particles 2 and 4 move inward. The conservation laws of charge, energy, linear momentum then read
\begin{equation}\label{charge_cons}
    q_1+q_2=q_3+q_4,
\end{equation}
\begin{equation}\label{kin_energy_cons}
    X_1+X_2=X_3+X_4,
\end{equation}
\begin{equation}\label{momentum_cons}
    Z_1-Z_2=Z_3-Z_4.
\end{equation}

Here, $X_i=E_i-q_i Q/r$, and $Z_i=\sqrt{X_i^2-m_i^2 f}$. We assume that the motion is purely radial, i.e., there is no angular momentum. We interpret $r$ as the radial coordinate of the collision process, which satisfies $r>r_+$, but can be arbitrarily close to the horizon if needed. We formulate the problem we are trying to achieve as follows: with fixed parameters of the colliding particles, is it possible for particle 3 to escape to infinity with arbitrarily large energy and/or mass?

As before, particle kinematics enforces the "forward-in-time" condition $X_i> 0$ (see sec. 2.3) for all $i=1,2,3,4$. In particular, it implies that $X_3< X_1+X_2$. We can thus state the following inequality:
\begin{equation}\label{charge_bound}
    q_3 \geq \frac{q_3 Q}{r_+}>\frac{q_3 Q}{r} = E_3-X_3>E_3-X_1-X_2
\end{equation}

Here we have used the fact that $Q\leq r_+$, which follows from the definition \eqref{roots}. This equation tells us that if we want to make $E_3$ arbitrarily large, it forces $q_3$ to have the same magnitude. In particular, energy extraction is impossible with electrically neutral particles ($q_3=0$), as for them \eqref{charge_bound} becomes $E_3<X_1+X_2$. This conclusion is valid for any collision process involving four particles in the RN metric.

If we are only interested in extracting energy, and not creating heavy particle, the problem admits a trivial solution. We can imagine two electrically neutral particles that collide and precisely exchange momentum, mass, and ``kinetic energy'' $X$, and the new particles have opposite charges of large absolute value:
\begin{equation}
    \begin{cases}
    q_1=q_2=0,\;q_3=-q_4=q,\\
    X_1=X_3,\; X_2=X_4,\\
    Z_1=Z_3,\; Z_2=Z_4,\\
    m_1=m_3,\; m_2=m_4.
    \end{cases}
\end{equation}

Clearly, this combination of parameters satisfies all conservation laws \eqref{charge_cons}--\eqref{momentum_cons}, as well as the forward-in-time conditions. The energy of particle 3 in this case is given by
\begin{equation}
    E_3=X_3+\frac{q_3 Q}{r}=X_1+\frac{qQ}{r}.
\end{equation}

Since $q$ is an arbitrary parameter, we can make it arbitrarily large in absolute value (with the same sign as $Q$). We thus see that in principle unbounded energy extraction is possible in this case. However, this fact is not particularly significant, because it does not even require curvature of spacetime. Moreover, the exact same process can be considered even in Newtonian mechanics with particles moving in Coulomb potential, giving the same results. Thus, it will prove to be more interesting to consider extraction of ultra-heavy particles, which will be shown to depend heavily on the curvature of spacetime.

We will assume that the collision occurs close to the event horizon, so that $f\ll 1$. We will assume that the particles 1, 2, 4 are ``usual" in the sense that for them $X_i\gg m_i f$, so $Z_i\approx X_i$ for $i=1,2,4$. We may not assume that this condition holds for particle 3, as will be clear from the result. Summing the energy and momentum conservation equations \eqref{kin_energy_cons}, \eqref{momentum_cons}) yields in this case
\begin{equation}
    2X_1=X_3+\sqrt{X_3^2-m_3^2 f},
\end{equation}
which allows us to solve for $X_3$:
\begin{equation}
    X_3=X_1+\frac{m_3^2 f}{4X_1}
\end{equation}

However, we have from $X_4>0$ that $X_3<X_1+X_2$. Therefore, we can deduce an upper bound on $m_3$:
\begin{equation}\label{upper_bound}
    m_3 < \sqrt{\frac{4X_1X_2}{f}}
\end{equation}

This inequality differs substantially from the inequalities such as \eqref{lower_bound} derived in~\cite{bib:zaslav3}, because it explicitly includes $f$, which is related to spacetime curvature. We claim that for a fixed $f$ it is possible to find a process that generates a particle of mass $m_3$ arbitrarily close to this upper bound. Therefore, if the collision occurs sufficiently close to the horizon (in other words, if $f$ is made sufficiently small), an ultra-massive particle detectable at infinity can be created.

\section{Other spacetimes}
\subsection{Majumdar-Papapetrou}
The Majumdar--Papapetrou spacetime describes the system of extremal black holes whose mutual gravitational attraction is canceled by their mutual electromagnetic repulsion. Here, we consider the system of two extremal black holes separated by a distance $2a$. The line element in Weyl's cylindrical coordinates $\{t\, \rho\, \varphi\, z\}$ has the following form:

\begin{equation} \label{metric_MP}
ds^2=-\frac{dt^2}{U^2}+U^2(d\rho^2+\rho^2d\phi^2+dz^2) \,.
\end{equation}

Where
\begin{equation}\label{U_mp}
U=1 + \frac{M_1}{\sqrt{\rho^2+(z-a)^2}} +\frac{M_2}{\sqrt{\rho^2+(z+a)^2}} \,.
\end{equation}
Here $M_1$ and $M_2$ are black hole masses which are equal to electric charges of black holes: $Q_1=M_1$ and $Q_2=M_2$.

The electric potential of \eqref{metric_MP}  is given by:
\begin{equation}
\varphi=1-\frac{1}{U}\,.
\end{equation}

One should note that the metric in Weyl's coordinates \eqref{metric_MP} describes the exterior of the black holes. The event horizons are collapsed into the points $\rho=0\,, z=\pm a$. The analogue description of a black hole like a point in Schwarzschild spacetime is described in~\cite{bib:dubrov}.

The Lagrangian of a massive charged test particle (with charge-to-mass ratio $\gamma$) is given by:
\begin{equation}
2\mathcal{L}=-U^{-2}\left( \frac{dt}{d\tau}\right)^2+U^2\left(\frac{d\rho}{d\tau}\right)^2+U^2\left(\frac{dz}{d\tau}\right)^2+\rho^2U^2\left(\frac{d\varphi}{d\tau}\right)^2-\gamma \varphi\frac{dt}{d\tau} \,.
\end{equation}

The MP spacetime doesn't depend on $t$ and $\varphi$ coordinates. This fact allows us to define two constants of motion measured by observer at infinity -- energy $E$ and projection of angular momentum $L_z$:
\begin{equation}
\begin{split}
E=U^{-2}\frac{dt}{d\tau}+\gamma \left(1-\frac{1}{U}\right) \,, \\
L_z=\rho^2U^2\frac{d\varphi}{d\tau} \,.
\end{split}
\end{equation}

We define $X=E-\gamma(1-\frac{1}{U})$, which plays a similar role in this metric as in the Reissner--Nordstr\"om one (see \eqref{def_xy}). For a particle moving along the $z$-axis, we can determine the components of 4-velocity from the normalization condition $g_{ik}u^iu^k=-1$:
\begin{equation} \label{eq:energ}
\begin{split}
u^0=\frac{dt}{d\tau}=U^2X \,, \\
u^1=\frac{d\rho}{d\tau}=0\,,\\
u^2 = \frac{d \phi}{d\tau}=0\,\\
u^3=\frac{dz}{d\tau}=\pm\sqrt{X^2-\frac{1}{U^2}} \,.
\end{split}
\end{equation}

Substituting \eqref{eq:energ} into  \eqref{ecm}, one obtains:

\begin{equation}
\frac{E^2_{c.m.}}{2m^2}-1=U^2\left(X_1X_2\pm  \sqrt{X_1^2-\frac{1}{U^2}}\sqrt{X_2^2-\frac{1}{U^2}}\right) \,,
\end{equation}
where the sign depends on the relative directions of the two $z$-axis velocities.

This analysis shows that BSW effect is possible in MP metric and produces certain restrictions on the particles, which precisely coincide with the conditions obtained for the RN metric.

It is reasonable to expect that MP near each the horizons should be equivalent to extremal RN metric. We have already derived the expansions for the RN case. Now we analyse near-horizon behaviour of MP near $z=+a$. For that, introduce a parameter $\delta_{MP}=(z-a)$. We demonstrate the equivalence along the z-axis by deriving expressions for the metric coefficients, electrostatic potential, and X.
\begin{itemize}
    \item Metric. The MP value corresponding to $f$ is $U^{-2}$. We can express it explicitly
    \bearr
        U^{-2}=\frac{(\delta(2a+\delta))^2}{(\delta(2a+\delta)+M_1(2a+\delta)+M_2\delta)^2}
        \nnn
        \approx \frac{4a^2}{4a^2M_1^2}\delta^2=\frac{\delta^2}{M_1^2}
    \ear
    Just as in \ref{f_crit_taylor} with $M_1=r_+$ identified.
    \item Potential. In extremal RN:
    \begin{equation}
        \phi = \frac{Q}{r}=\left(1+\frac{\delta}{M}\right)^{-1}\approx 1-\frac{\delta}{M}+\frac{\delta^2}{M^2}+...
    \end{equation}
    In MP:
    \bearr
        \phi = \frac{U-1}{U}
        =\left(1+\frac{M_1+M_2}{2aM_1}\delta\right)
        \nnn
        \times\left(1+\frac{M_1+M_2+2a}{2aM_1}\delta+\frac{1}{2aM_1}\delta^2\right)^{-1}
        \nnn
        =1-\frac{\delta}{M_1}+\frac{M_2+2a}{2aM_1^2}\delta^2+...
    \ear
    Note that in case $a\to\infty$ or $M_2\to 0$  this reduces to the RN expression (either of these conditions makes the second black hole irrelevant to the neighbourhood of the first one).
    \begin{equation}
        X = E-\gamma\phi
    \end{equation}
    Then, $X=E-\gamma\phi$ is expanded trivially. Also,
    \begin{equation}
        X^2 = X_h^2+2X_h\frac{\gamma}{M_1}\delta+\left(\frac{\gamma^2}{M_1^2}-2X_h\gamma\frac{2a+M_2}{2aM_1}\right)\delta^2+...
    \end{equation}
    One can clearly see that the 2nd kinematic condition $X^2-Yf\geq 0$ (see sec. 2.3) leads to the same result as before for $L=0$. I.e. $|\gamma|>1$, and the 1st condition implies it is positive.
\end{itemize}
These results demonstrate definitively, that BSW effect for motion along z-axis in MP metric is exactly identical to radial motion in extremal RN.
\subsection{Bardeen regular black hole}
The first solution of the Einstein equations describing the regular black hole has been obtained by Bardeen~\cite{bib:bardeen}, which is given by~\cite{bib:ansoldy}
\begin{equation} \label{eq:bardeen}
\begin{split}
ds^2=-f(r)dt^2+f(r)^{-1} dr^2+r^2d\Omega^2 \,, \\
f(r)\equiv 1-\frac{2Mr^2}{(r^2+g^2)^{\frac{3}{2}}}\,.
\end{split}
\end{equation}
Here $d\Omega^2=d\theta^2+\sin^2\theta d\varphi^2$ is the geometry on the unit two-sphere, $M$ is the mass of a black hole, $g$ is the magnetic monopole charge of the non-linear self gravitating magnetic field. 
The solution \eqref{eq:bardeen} is interpreted as the gravitational field of a nonlinear monopole, i.e., as a magnetic solution to Einstein field equations coupled to a non-linear electrodynamics. 
Our consideration will go close to ~\cite{bib:zaslav3} but with several additions:
\begin{itemize}
\item first of all, we don't assume the black hole to be critical one i.e. $f'(r_h)\neq 0$ (Where $r_h$ is the event horizon location);
\item we don't specify the potential $\phi$ of the external magnetic field.
\end{itemize}

The BSW effect has been considered for Bardeen and other regular black holes in the paper~\cite{bib:regularbsw}. 

We consider the near horizon collision of two particles $1$ and $2$ which have been injected from infinity. In the process of collision we have particle $3$ which goes away from black hole and a particle $4$ which falls into a black hole. We don't consider the mass of particles, assuming that they are equall - it is not important for the analysis below. Also, we assume radial motion of the particles i.e. $L_i=0 \,, i= 1, 2, 3, 4$, where $L$ is an angular momentum per unit mass.
The conservation of magnetic charge $\mu$, energy per unit mass $E$ and radial momentum give us
\begin{equation}
\mu_1+\mu_2=\mu_3+\mu_4 \,.
\end{equation}
\begin{equation}
\begin{split} \label{eq:benergy}
X_1+X_2=X_3+X_4 \,, \\
X_i\equiv E_i-\mu_i\phi  \,.
\end{split}
\end{equation}
\begin{equation}
\begin{split} \label{eq:bradial}
-Z_1-Z_2=Z_3-Z_4 \,, \\
Z_i=\sqrt{X_i^2-f(r)} \,.
\end{split}
\end{equation}

Following the classification, given in~\cite{bib:zaslav2}, we consider three types of particles:
\begin{enumerate}
\item Usual particle. We consider near horizon collision where $f(r)\ll 1$ which gives us
\begin{equation} \label{eq:busual}
z\approx X_h-\frac{f}{2X_h} \,.
\end{equation}
By subscript $h$ we denote quantites evaluated on the event horizon.
\item Critical particle. For critical particle $X_h=0$. In our case for critical particle we put $z=0$.
\item Slightly non-critical particle. In this case we assume
\begin{equation} \label{eq:bassume}
\mu=\frac{E(1-\delta)}{\phi} \,, \delta>0 \,.
\end{equation}
In this case, one can write:
\begin{equation} \label{eq:bslightly}
z\approx E\delta-\frac{f}{2E\delta} \,.
\end{equation}
\end{enumerate}

We assume that particles $1$ and $4$ are usual particles, the particle $2$ is critical and particle $3$ is slightly non-critical.

Substituting \eqref{eq:busual}, \eqref{eq:bassume}, \eqref{eq:bslightly} into \eqref{eq:bradial} and using \eqref{eq:benergy} to eliminate $X_4$ ($X_4=X_1-X_3$), one obtains the following expression: 
\begin{equation} \label{eq:fdsa}
\frac{f}{2X_{h,1}}=2E_3\delta -\frac{f}{2E_3\delta}+\frac{f}{2X_{h,1}-2E_3\delta} \,.
\end{equation}
Reducing the expression \eqref{eq:fdsa} to a common denominator (We multiply \eqref{eq:fdsa} by $8X_{h,1}E_3\delta(X_{h,1}-E_3\delta)$) and neglecting $O(\delta^3)$, one obtains the following inequality: 

\begin{equation}
\begin{split}
\alpha E_3^2+X_{h,1}\delta f E_3-X^2_{h,1}f \geq 0 \,, \\
\alpha \equiv 4X^2_{h,1}\delta^2+\delta^2f >0 \,.
\end{split}
\end{equation}

From this inequality we obtain lower restriction on $E_3$:
\begin{equation}
E_3\geq \frac{X_{h,1}\delta f+\sqrt{X^2_{h,1}\delta^2f^2+4X^2_{h,1}\alpha}}{2\alpha} \,.
\end{equation}

Like in~\cite{bib:zaslav3} we didn't find upper limit for $E_3$. However, as we have shown in sec. 2.5, in RN spacetime, energy extraction for real particles is negligibly small. We expect this to be true for Bardeen metric as well.

\section{Conclusion}

The conditions for divergent center-of-mass energy in the RN metric and in the MP metric are found to be very strict: the work suggests conditions on the criticality of the particles and extremality on the black holes. It seems that satisfaction of these conditions is very unlikely. There is a possibility that the used models are not fully encapsulating all the important features of real black holes (for example, there was no mention of the accretion disk). Therefore, the obtained conditions could be inapplicable to real black holes so that high-energy collisions are more likely. Therefore, it may be beneficial to perform a similar analysis for a wider range of black hole models, searching for a metric, where such conditions would be more realistic, so that these effects would have a chance of being observed experimentally.

We have shown that energy extraction is possible also in Bardeen regular black hole spacetime. Like in usual Reissner-Nordstrom spacetime~\cite{bib:zaslav3}, we were able to find only the lower restriction for the extracted energy. However, for RN black hole, we have shown in Table \eqref{tab:real_particles} that energy extraction in collisions of real particles falling into black holes have extremely low lower bounds on extracted masses: even for relatively heavy particles like the nucleus of gold atom it is many orders of magnitude lower than the mass of electron, so it only tells us that these processes cannot emit ultra-light particles like neutrinos. The same results, we expect in Bardeen regular black hole despite that the upper bound is absent. It is reasonable to assume that even if collisions with creation of ultra-heavy particles escaping to infinity are physically possible, they must be severely outnumbered by collisions that produce usual particles with masses smaller than the masses of collision's constituents. Therefore, this result (existence of lower bound) is probably not useful for actual astrophysical purposes.

On the other hand, the derived inequality \eqref{upper_bound} differs substantially from the inequalities such as \eqref{lower_bound} derived in~\cite{bib:zaslav3}, because it explicitly includes $f$, which is related to the spacetime curvature. We claim that for a fixed $f$ it is possible to find a process that generates a particle of mass $m_3$ arbitrarily close to this upper bound. Therefore, if the collision occurs sufficiently close to the horizon (in other words, if $f$ is made sufficiently small), an ultra-massive particle detectable at infinity can be created. However, in derivation we avoided the problem of existence of particles moving outward in close vicinity of the event horizon. It deserves a further investigation to understand if such particles actually exist near black holes in our universe.

\Acknow{We thank Dr. Oleg Zaslavskii (Kharkiv National University) for comments and suggestions.}

\Funding{We were funded by grant NUM. 22-22-00112 RSF, and supported by Letovo School Charity Fund.}


\end{document}